\documentclass[aps,prd,twocolumn,superscriptaddress,nofootinbib]{revtex4-1}
\usepackage{graphicx}
\usepackage[colorlinks,citecolor=blue]{hyperref}
\usepackage{amsmath,amssymb,latexsym,bm,braket}
\usepackage{hepunits}
\usepackage{slashed}
\usepackage{color}
\usepackage[utf8]{inputenc}

\newcommand{\met}{\slashed{\rm E}_{T}}

\allowdisplaybreaks

\begin{document}

\title{One Bump or Two Peaks? The 750 GeV Diphoton Excess and Dark Matter with a Complex Mediator}

\author{Qing-Hong Cao}
\email{qinghongcao@pku.edu.cn}
\affiliation{School of Physics and State Key Laboratory of Nuclear Physics and Technology, Peking University, Beijing 100871, China}
\affiliation{Collaborative Innovation Center of Quantum Matter, Beijing, China}
\affiliation{Center for High Energy Physics, Peking University, Beijing 100871, China}

\author{Yin-Qiang Gong}
\email{gongyq@pku.edu.cn}
\affiliation{School of Physics and State Key Laboratory of Nuclear Physics and Technology, Peking University, Beijing 100871, China}

\author{Xing Wang}
\email{x.wong@pku.edu.cn}
\affiliation{School of Physics and State Key Laboratory of Nuclear Physics and Technology, Peking University, Beijing 100871, China}

\author{Bin Yan}
\email{binyan@pku.edu.cn}
\affiliation{School of Physics and State Key Laboratory of Nuclear Physics and Technology, Peking University, Beijing 100871, China}

\author{Li Lin Yang}
\email{yanglilin@pku.edu.cn}
\affiliation{School of Physics and State Key Laboratory of Nuclear Physics and Technology, Peking University, Beijing 100871, China}
\affiliation{Collaborative Innovation Center of Quantum Matter, Beijing, China}
\affiliation{Center for High Energy Physics, Peking University, Beijing 100871, China}

\begin{abstract}
We consider the possibility that the recently observed excess in the diphoton invariant mass spectrum around 750 GeV is due to two narrow scalars. We demonstrate that there is no need to introduce invisible decay modes to enlarge the widths as long as the two scalars exhibit an appropriate mass splitting. Nevertheless we investigate the possible connection of these two scalars to the dark matter in a toy model with a Majorana fermionic dark matter candidate. We explore how large the widths of the two scalar widths could be taking into account various constraints from collider and dark matter experiments. Introducing two scalars alleviates the tension between the diphoton production and dark matter constraints. 
\end{abstract}

\maketitle

\section{Introduction}

Recently, some excess in the diphoton invariant mass spectrum is observed around 750~GeV by both the ATLAS and the CMS collaborations at the LHC Run-II~\cite{ATLAS:2015xxx,CMS:2015dxe}.
The large excess around 750~GeV still remains in the recently updated analysis by the ATLAS and CMS~\cite{ATLAS-CONF-2016-018, CMS-PAS-EXO-16-018}. The diphoton invariant mass spectrum in particular does not change much.
It has drawn a lot of attentions in the field of particle physics~\cite{Mambrini:2015wyu,Backovic:2015fnp,Angelescu:2015uiz,Nakai:2015ptz,Knapen:2015dap,
Harigaya:2015ezk,Buttazzo:2015txu,Pilaftsis:2015ycr,Franceschini:2015kwy,
DiChiara:2015vdm,Higaki:2015jag,McDermott:2015sck,Ellis:2015oso, Low:2015qep,Bellazzini:2015nxw,Gupta:2015zzs,Petersson:2015mkr,Molinaro:2015cwg,Dutta:2015wqh,
Cao:2015pto,Matsuzaki:2015che,Kobakhidze:2015ldh,Cox:2015ckc,Becirevic:2015fmu,No:2015bsn,Demidov:2015zqn,
Chao:2015ttq,Fichet:2015vvy,Curtin:2015jcv,Bian:2015kjt,Chakrabortty:2015hff,Ahmed:2015uqt,Agrawal:2015dbf,
Csaki:2015vek,Falkowski:2015swt,Aloni:2015mxa,Bai:2015nbs,Gabrielli:2015dhk,
Benbrik:2015fyz,Kim:2015ron,Alves:2015jgx,
Megias:2015ory,Bernon:2015abk,Chao:2015nsm,Arun:2015ubr,Han:2015cty,Chang:2015bzc,Chakraborty:2015jvs,
Ding:2015rxx,Han:2015dlp,Han:2015qqj,Luo:2015yio,Chang:2015sdy,Bardhan:2015hcr,Feng:2015wil,Antipin:2015kgh,
Wang:2015kuj,Cao:2015twy,Huang:2015evq,Liao:2015tow,Heckman:2015kqk,Dhuria:2015ufo,Bi:2015uqd,Berthier:2015vbb,
Cho:2015nxy,Cline:2015msi,Chala:2015cev,Barducci:2015gtd,Boucenna:2015pav,Murphy:2015kag,
Hernandez:2015ywg,Dey:2015bur,Pelaggi:2015knk,deBlas:2015hlv,Belyaev:2015hgo,Huang:2015rkj,
Moretti:2015pbj,Patel:2015ulo,Badziak:2015zez,Chakraborty:2015gyj,Cao:2015xjz,
Altmannshofer:2015xfo,Cvetic:2015vit,Gu:2015lxj,Allanach:2015ixl,Davoudiasl:2015cuo,
Craig:2015lra,Das:2015enc,Cheung:2015cug,Liu:2015yec,Zhang:2015uuo,Casas:2015blx,Han:2015yjk,Li:2015jwd,
Son:2015vfl,Tang:2015eko,An:2015cgp,Cao:2015apa,Wang:2015omi,Cai:2015hzc,Cao:2015scs,Gao:2015igz,Bi:2015lcf,Chao:2015nac,
Goertz:2015nkp,Dev:2015vjd,Bizot:2015qqo,Ibanez:2015uok,
Kang:2015roj,Hamada:2015skp,Huang:2015svl,Kanemura:2015bli,Kanemura:2015vcb,Low:2015qho,
Hernandez:2015hrt,Jiang:2015oms,Kaneta:2015qpf,Dasgupta:2015pbr,Jung:2015etr,
Palti:2016kew,Nomura:2016fzs,Han:2016bus,Ko:2016lai,Chao:2016mtn,
Karozas:2016hcp,Hernandez:2016rbi,Dutta:2016jqn,Deppisch:2016scs,Ito:2016zkz,Zhang:2016pip,Berlin:2016hqw,
Bhattacharya:2016lyg,Sahin:2016lda,Fichet:2016pvq,Borah:2016uoi,Stolarski:2016dpa,Hati:2016thk,Ko:2016wce,
Cao:2016udb,Ding:2016ldt,Davis:2016hlw,Faraggi:2016xnm,Dorsner:2016ypw,Nomura:2016seu,Chao:2016aer,Han:2016bvl,
Franzosi:2016wtl}. In these works, the excess is usually interpreted as due to a new scalar particle arising in some new physics model. In the current experimental data, the excess appears in a wide range of the diphoton invariant mass. As a result, the corresponding scalar particle seems to exhibit a large width, which is hard to understand in various new physics models. An intriguing assumption is that the 750~GeV scalar predominantly decays into a pair of invisible dark matter candidates, yielding a large width~\cite{Backovic:2015fnp}. The large partial width of the dark matter decay mode, however, has a tension with the current experimental data on dark matter detection~\cite{Bi:2015uqd}. Another way out of the predicament is to interpret the broad peak as generated by the kinematics effects, e.g. the three body decays~\cite{Bernon:2015abk}, four body decays~\cite{Cho:2015nxy}, or multiple-photon decays~\cite{Franceschini:2015kwy,Huang:2015evq,Chang:2015sdy}, etc. 

In this work we consider an alternative origin of the broad excess around 750~GeV. 
In the early operation of the LHC Run-II the photon energy resolution is about 6--10~GeV, which leads to a large uncertainty in the diphoton invariant mass. The available integrated luminosity is also very limited and the bin sizes of the experimental data are rather large. As a result, one cannot distinguish a wide scalar resonance from two narrower scalar resonances with a certain level of overlap within the current experimental accuracy~\cite{Franceschini:2015kwy,Petersson:2015mkr}. 
Such nearly degenerate scalar resonances can arise naturally in some supersymmetric models~\cite{Petersson:2015mkr,Casas:2015blx,Wang:2015omi,Ding:2016udc}.
Since the two scalars can be narrow, one does not need to introduce the dark matter decay mode to enlarge the width, and the tension between the broad width and dark matter experiments are greatly eliminated. Of course, it is too early to draw any affirmative conclusion on whether the broad excess is due to one wide resonance or two narrow resonances, or even whether the excess is genuine at all. However, it is still instructive to consider the possibility of the 750~GeV resonance being a composite of two scalars. Four interesting questions are addressed below: {\it i}) could two narrow scalars explain the broad resonance around 750~GeV? {\it ii}) how narrow could the two scalars be? {\it iii}) how wide could the two scalars be if both the scalars talk to dark matter candidates? {\it iv}) could the two scalars be probed by the collider and non-collider experiments in the near future?

The paper is organized as follows. In Sec.~II we introduce our toy model of a complex scalar and discuss its relation to the diphoton excess. In Sec.~III we couple this complex scalar to a dark matter candidate and examine the constraints coming from dark matter experiments. Finally, we conclude in Sec.~IV.

\section{Toy model}

In order to mimic the two-peak structure around 750~GeV, we consider a toy model which consists of a complex scalar field $\Phi = (S + iA)/\sqrt{2}$ in the singlet representation of the standard model (SM) gauge group. We assume that no additional source of CP violation arises from this new scalar sector. The most general mass term for this complex scalar can therefore be written as 
\begin{align}
  \mathcal{L}_{\text{mass}} &= - M^2 \left| \Phi \right|^2 - \frac{1}{2} m^2 \left( \Phi^2 + \Phi^{*2} \right)\nonumber\\
  & = - \frac{1}{2} \left( M^2 + m^2 \right) S^2 - \frac{1}{2} \left( M^2 - m^2 \right) A^2 \, ,
\end{align}
from which it is clear that a relatively small mass splitting between the real and imaginary parts can be achieved without too much tuning. The masses of the scalar $S$ and the pseudoscalar $A$ are given by
\begin{align}
  M_S &= \sqrt{M^2+m^2} = M + \Delta M \, , \nonumber \\
  M_A &= \sqrt{M^2-m^2} = M - \Delta M \, ,  
\end{align}
where $\Delta M \approx m^2/(2M)$. The complex scalar can couple to SM gauge bosons at one-loop with additional new particles (scalars, fermions or vector bosons) running in the loop. In the limit when the particles in the loop are much heavier than the complex scalar, these interactions can be represented by an effective Lagrangian
\begin{align}
  \mathcal{L}_{\text{eff}} &= \frac{c_g^S}{M} S G^a_{\mu\nu} G^{a\mu\nu} + \frac{c_W^S}{M} S W^a_{\mu\nu} W^{a\mu\nu} + \frac{c_B^S}{M} S B_{\mu\nu} B^{\mu\nu} \nonumber
  \\
  &+ \frac{c_g^A}{M} A G^a_{\mu\nu} \tilde{G}^{a\mu\nu} + \frac{c_W^A}{M} A W^a_{\mu\nu} \tilde{W}^{a\mu\nu} + \frac{c_B^A}{M} A B_{\mu\nu} \tilde{B}^{\mu\nu} \, , \nonumber
\end{align}
where the dual field strength tensors are defined by, e.g., $\tilde{G}^a_{\mu\nu} = \epsilon_{\mu\nu\alpha\beta} G^{a\alpha\beta} / 2$. Through these effective operators, $S$ and $A$ can decay into $W^+W^-$, $ZZ$ and $Z\gamma$ in addition to $\gamma\gamma$ and $gg$. For the moment we concentrate on the following part of the effective Lagrangian relevant for diphoton production:
\begin{align}
  \mathcal{L}_{\text{eff}} &\supset \frac{c_g^S}{M} S G^a_{\mu\nu} G^{a\mu\nu} + \frac{c_\gamma^S}{M} S F_{\mu\nu} F^{\mu\nu} \nonumber\\ 
  &+ \frac{c_g^A}{M} A G^a_{\mu\nu} \tilde{G}^{a\mu\nu} + \frac{c_\gamma^A}{M} A F_{\mu\nu} \tilde{F}^{\mu\nu} \, ,
\end{align}
where $c_\gamma^{S/A} = c_B^{S/A} \cos^2\theta_W + c_W^{S/A} \sin^2\theta_W$ with $\theta_W$ the weak mixing angle.

The differential cross section for the production of a pair of photons through the scalar resonance ($gg \to S \to \gamma\gamma$) can be written as
\begin{align}
  \frac{d\sigma}{dQ^2} = \frac{\pi \Gamma^S_{gg} \Gamma^S_{\gamma\gamma}}{8s} \, \frac{M_{\gamma\gamma}^6}{M_S^6} \, \frac{\mathcal{L}_{gg}(Q^2/s) \, K_S}{(M_{\gamma\gamma}^2-M_S^2)^2 + M_S^2 \Gamma_S^2} \,  ,
\end{align}
where $\Gamma_S$ is the total width of the scalar, $M_{\gamma\gamma}$ is the invariant mass of the photon pair, $K_S$ is the $K$-factor arising from higher order QCD corrections, the partial decay widths of $S$ are given by
\begin{align}
& \Gamma^S_{gg} \equiv \Gamma(S \to gg) = \left( \frac{c_g^S}{M} \right)^2 \frac{2M_S^3}{\pi} \, ,\nonumber\\ 
& \Gamma^S_{\gamma\gamma} \equiv \Gamma(S \to \gamma\gamma) = \left( \frac{c_\gamma^S}{M} \right)^2 \frac{M_S^3}{4\pi} \, ,
\end{align}
and the gluon luminosity function is defined as
\begin{align}
  \mathcal{L}_{gg}(\tau) = \int_\tau^1 \frac{dx}{x} f_g(x) f_g(\tau/x) \, .
\end{align}
In the narrow width limit $\Gamma_S \ll M_S$, the total cross section can be approximated as
\begin{align}
  \sigma(pp \to S \to \gamma\gamma) \approx \frac{\Gamma^S_{gg} \Gamma^S_{\gamma\gamma}}{s M_S \Gamma_S} \, \frac{\pi^2}{8} \mathcal{L}_{gg}(M_S^2/s) \, K_S \, .
\end{align}
The cross section with the pseudoscalar resonance can be obtained from the above formulas by obvious substitutions. 
Note that the two channels do not interfere due to different CP properties, and the total rate is simply the sum of the two contributions.
In the following, we are going to make further simplifications by assuming
\begin{align}
\label{eq:assumptions1}
& \Gamma^S = \Gamma^A = \Gamma \, , \quad &&\Gamma^S_{gg} = \Gamma^A_{gg} = \Gamma_{gg} \, , \nonumber\\
& \Gamma^S_{\gamma\gamma} = \Gamma^A_{\gamma\gamma} = \Gamma_{\gamma\gamma} \, , \quad &&K_S = K_A = K \, .
\end{align}
The total diphoton production rate through the complex mediator can then be approximated by
\begin{align}
  \sigma_{\gamma\gamma} = \sigma(pp \to \Phi \to \gamma\gamma) \approx 2 \times \frac{\Gamma_{gg}\Gamma_{\gamma\gamma}}{s M \Gamma} \, \frac{\pi^2}{8} \mathcal{L}_{gg} (M^2/s) \, K \, ,\nonumber
\end{align}
where we have used $M_S \approx M_A \approx M$ and the factor of 2 accounts for the contributions from the two resonances. Using $M \approx \unit{750}{\GeV}$, we obtained the following relation between the total width and the two partial widths
\begin{align}
  \label{eq:Gamma}
\Gamma \approx \frac{\unit{10}{\femtobarn}}{\sigma_{\gamma\gamma}} \, \Gamma_{gg} \Gamma_{\gamma\gamma} \times \unit{1950}{\GinveV} \, ,
\end{align}
where we have used $K \approx 1.48$. On the other hand, the null result of the searches for dijet resonances at \unit{8}{\TeV} imposes a constraint $\Gamma_{gg}^2 / \Gamma < \unit{0.6}{\GeV}$. Therefore, if only the $gg$ and $\gamma\gamma$ decay modes are present and we also assume $\Gamma_{gg} \sim \Gamma_{\gamma\gamma}$ (which is natural in most UV completed models), the total width cannot exceed $\mathcal{O}(\unit{1}{\GeV})$.  Adding the $W^+W^-$, $ZZ$ and $Z\gamma$ decay modes do not change this qualitative conclusion taking into account current experimental constraints. Note that if only a single resonance is present, the observed excess in a broad range of $M_{\gamma\gamma}$ does not favor such a small total width. However, with two resonances, a narrow width is allowed due to the mass splitting and the limited statistics of the experimental data.

\begin{figure}
\includegraphics[width=0.6\linewidth]{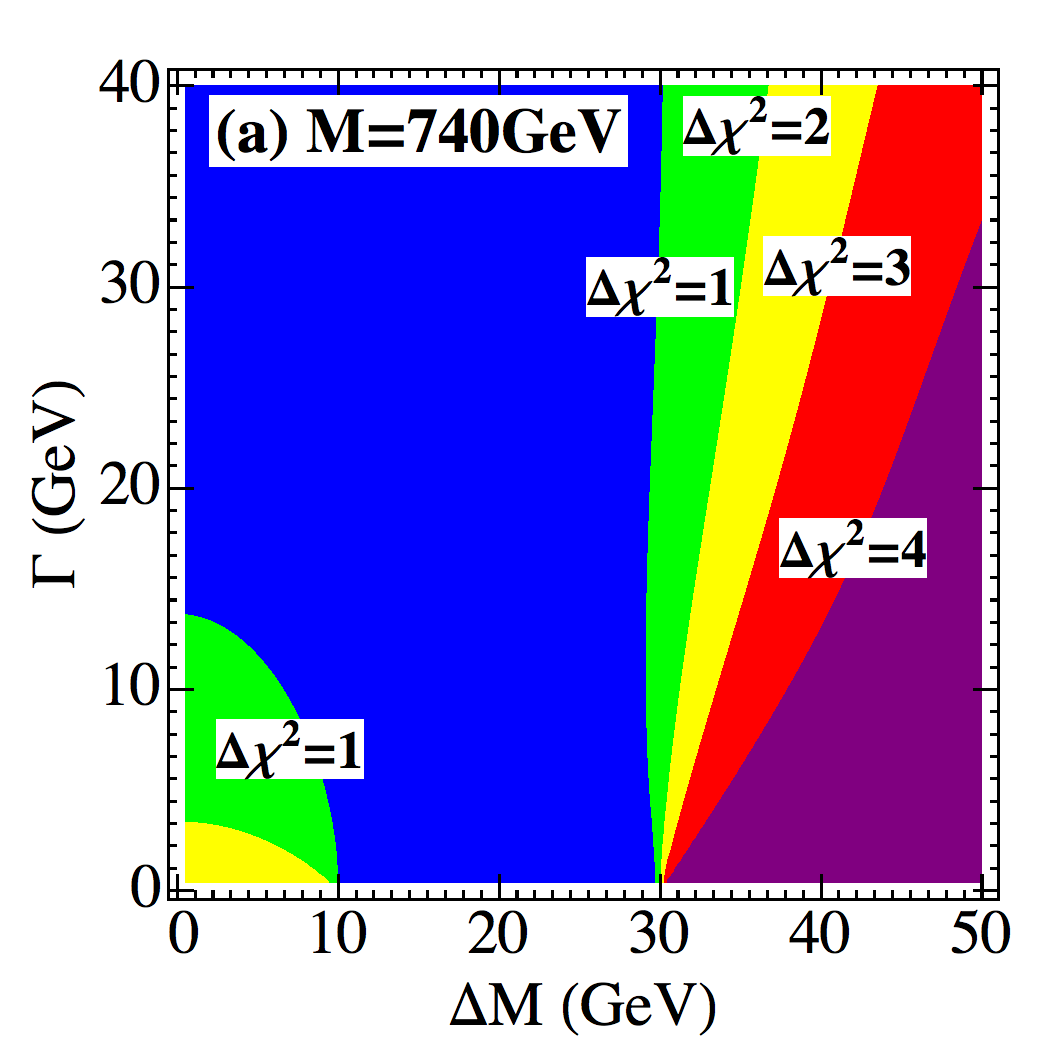}
\includegraphics[width=0.6\linewidth]{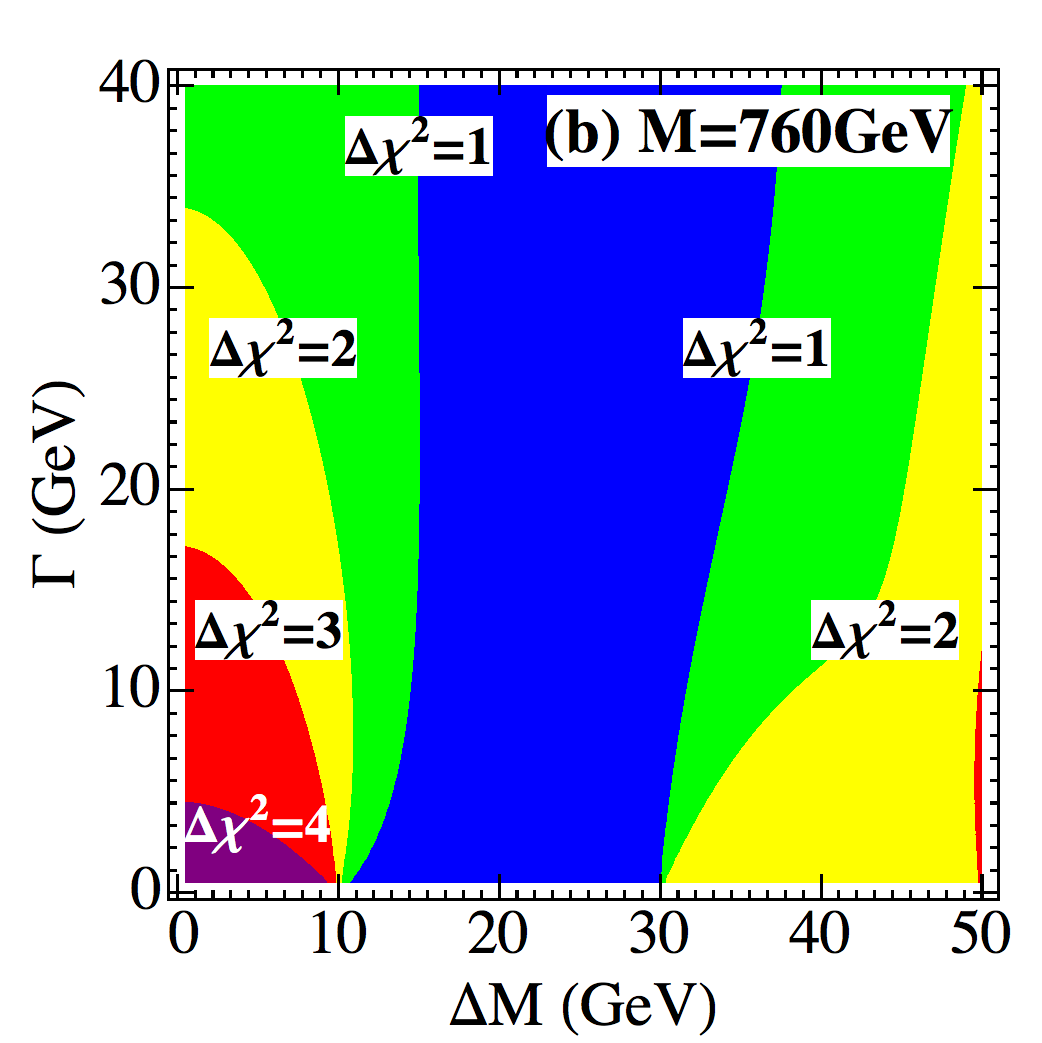}
\caption{Contour plots of $\Delta \chi^2$ as a function of $\Delta M$ and $\Gamma$ with $M=\unit{740}{\GeV}$ (a) and $M=\unit{760}{\GeV}$ (b).}
\label{fig:chi2}
\end{figure}

We now perform a $\chi^2$-fit to the ATLAS data in the 4 bins around \unit{750}{\GeV} allowing the total width $\Gamma$ to vary independently. The $\chi^2$ is defined as
\begin{align}
  \chi^2 = \sum_{i=1}^4 \frac{\left( N_i^{\text{obs}} - N_i^{\text{th}} \right)^2}{\sigma_i^2} \, .
\end{align}
In the above formula, $N_i^{\text{obs}}$ and $N_i^{\text{th}}$ are the observed and the theoretically expected number of events in the $i$-th bin, respectively, and $\sigma_i$ is the corresponding variance. The theoretically expected numbers $N_i^{\text{th}}$ depend on the model parameters $\{M, \Delta M, \Gamma, \Gamma_{gg}\Gamma_{\gamma\gamma}\}$, as well as on the acceptance times efficiency $A \times \epsilon$. For the latter we assume a constant value $A \times \epsilon = 0.4$. In order to present the fit results more clearly, we fix the value of $M$ to be \unit{740}{\GeV} or \unit{760}{\GeV}, and for each pair of values of $\Delta M$ and $\Gamma$, we fix the value of $\Gamma_{gg}\Gamma_{\gamma\gamma}$ by requiring that the corresponding $\chi^2$ is minimized. We vary $\Delta M$ and $\Gamma$ in the range $0 < \Delta M < \unit{50}{\GeV}$ and $0 < \Gamma < \unit{40}{\GeV}$ to find the global minimum $\chi^2_{\text{min}}$. The resulting $\Delta\chi^2 = \chi^2 - \chi^2_{\text{min}}$ as a function of $\Delta M$ and $\Gamma$ is shown as contour plots in Fig.~\ref{fig:chi2}.

\begin{figure}
\includegraphics[scale=0.3]{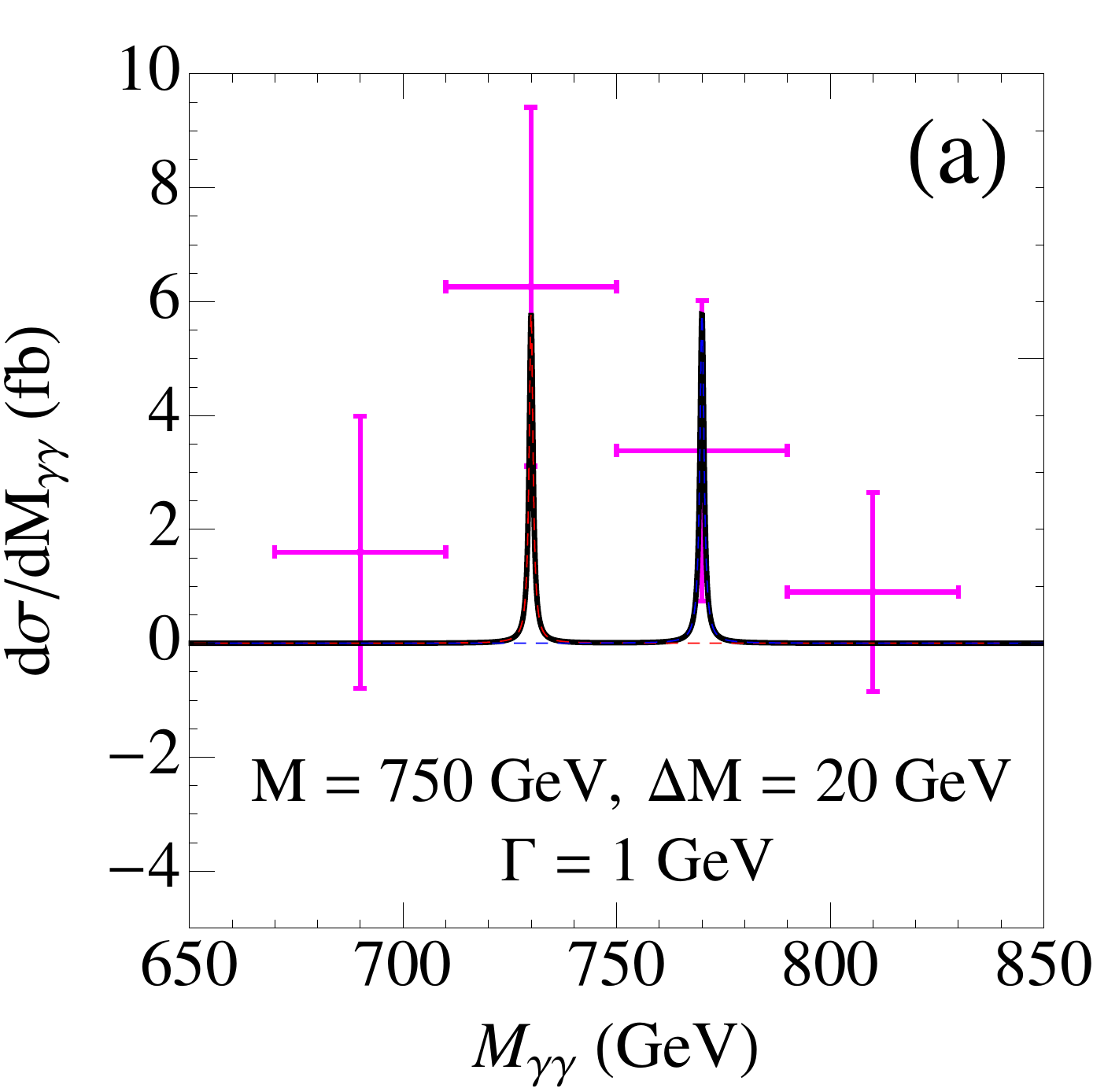}
\includegraphics[scale=0.3]{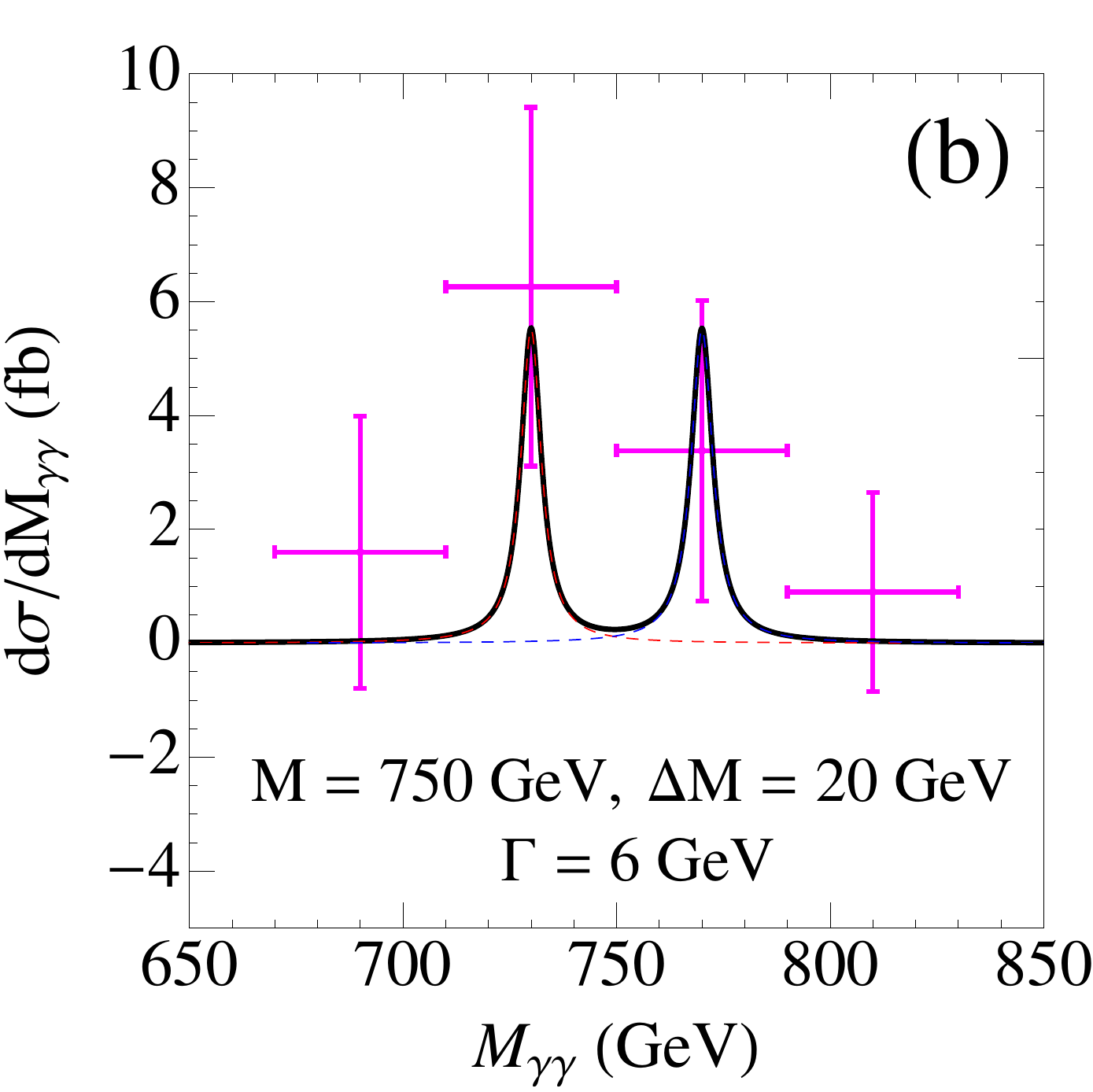}\\
\includegraphics[scale=0.3]{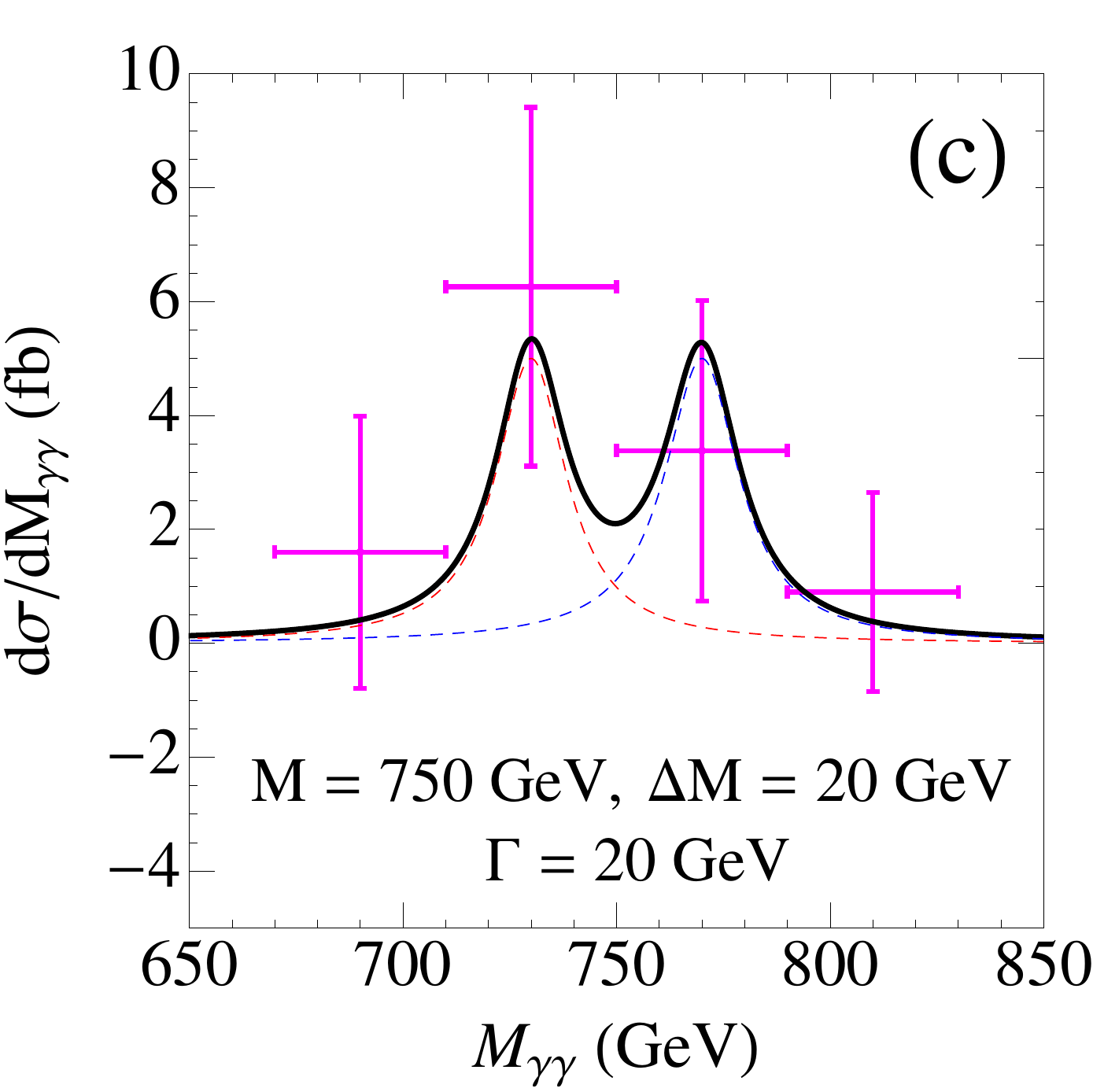}
\includegraphics[scale=0.3]{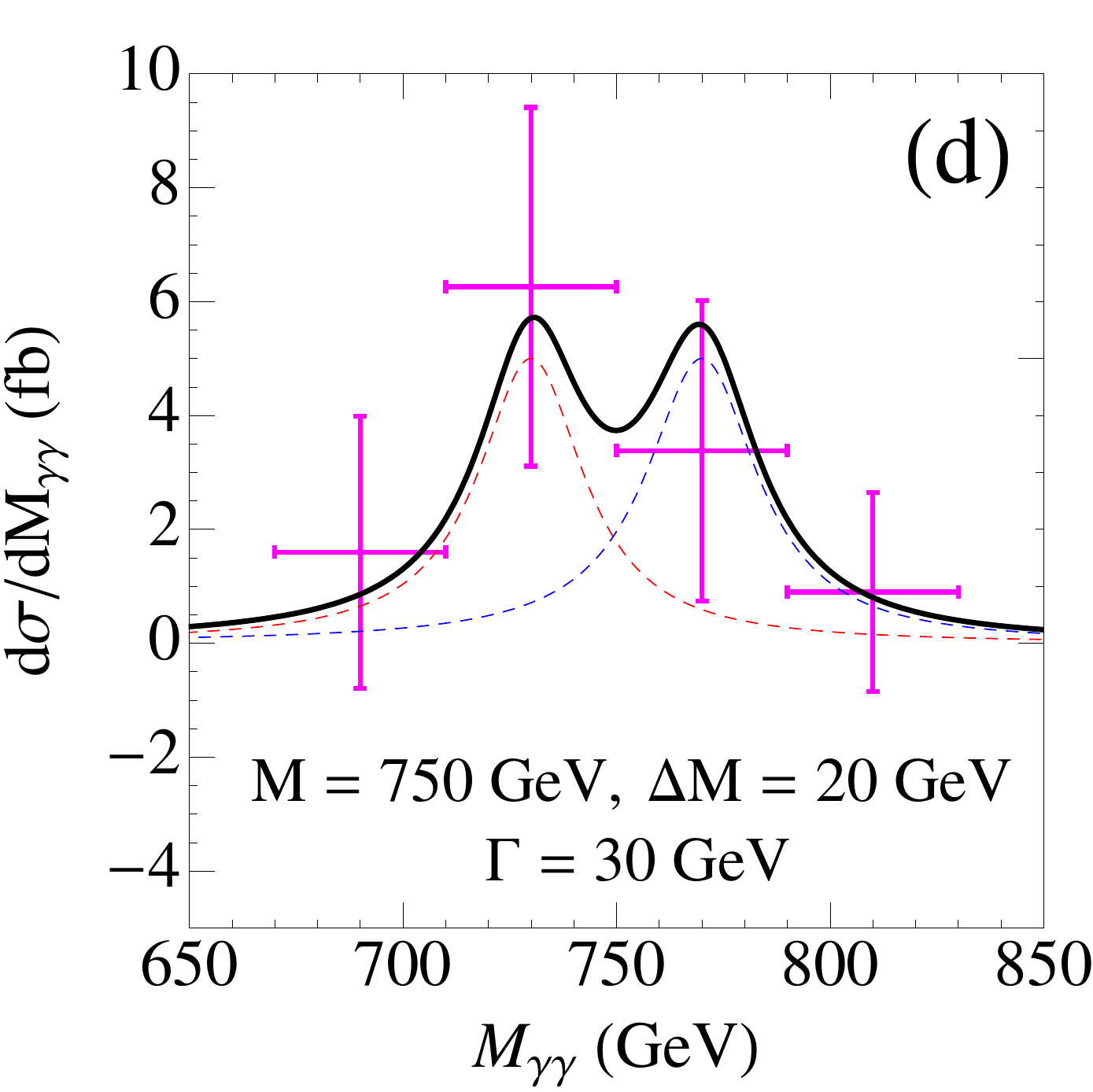}\\
\includegraphics[scale=0.3]{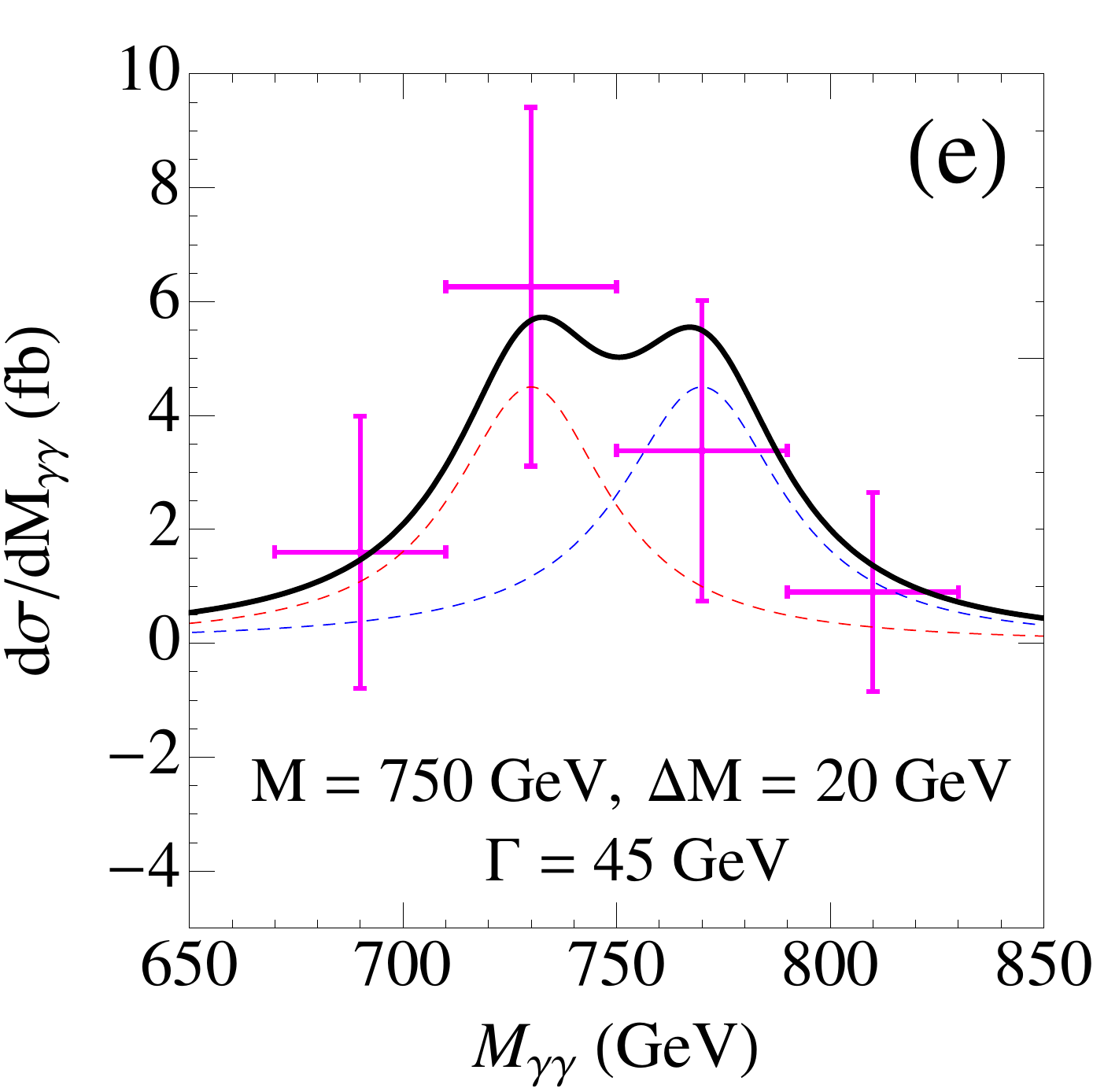}
\includegraphics[scale=0.3]{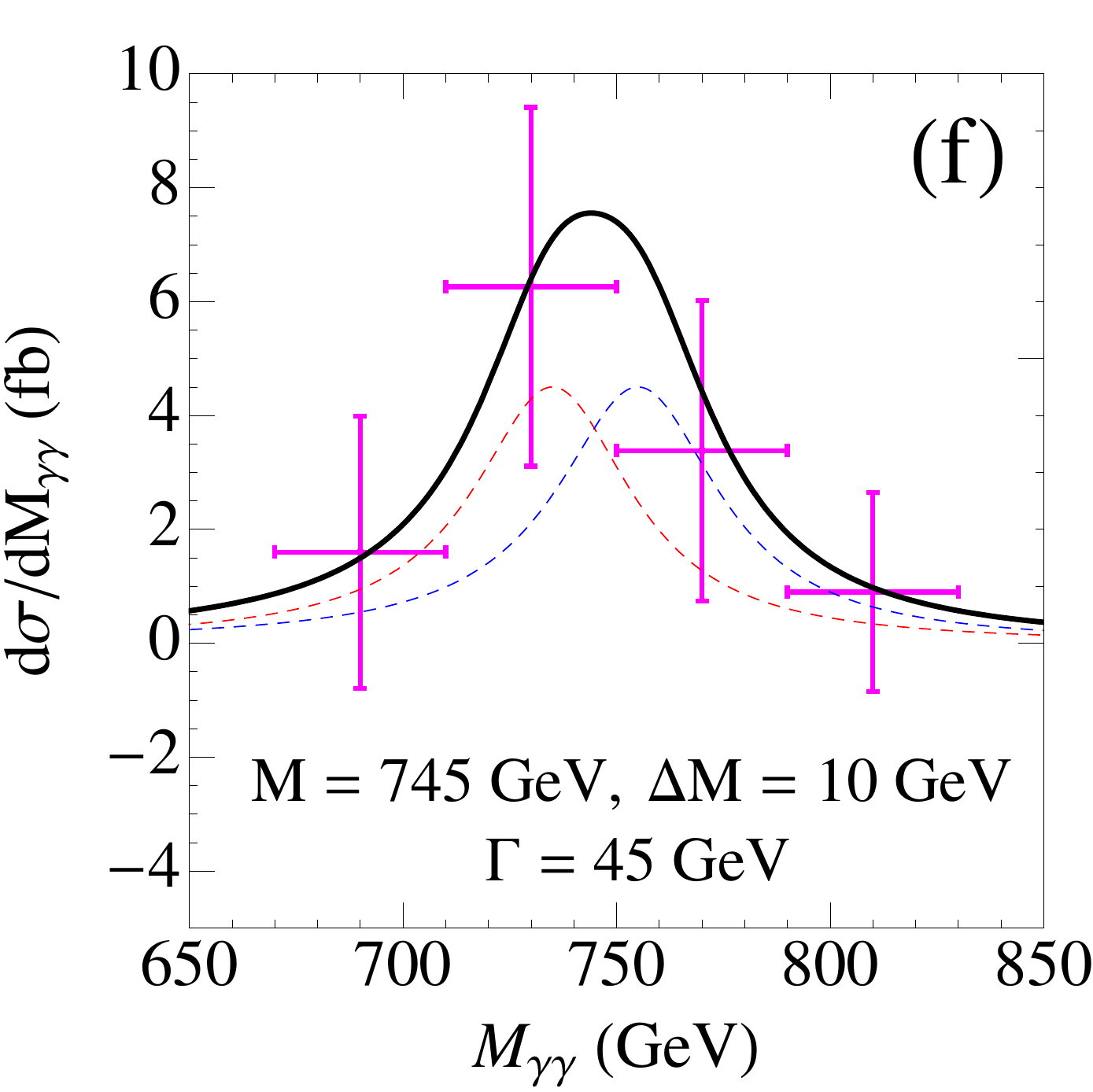}
\caption{Diphoton invariant mass distributions where the red and blue dashed curves represent each individual scalar contribution and the black curves denote their sums. The top row shows the narrow scalars while the bottom row displays the broad scalars. }
\label{fig:fitting}
\end{figure}

There are a few messages one can read off from the plots in Fig.~\ref{fig:chi2}. First of all, with a small mass splitting which is effectively equivalent to a single resonance, a narrow width of a few GeV is indeed not favored by the experimental data, although it is not excluded either. On the other hand, the quality of the fit mildly depends on $\Delta M$ and $\Gamma$, as long as the two peaks separately reside in the two central bins around \unit{750}{\GeV}. This can be expected from the large size of the bins due to the low statistics. For illustration, in Fig.~\ref{fig:fitting} we show the experimental data on $M_{\gamma\gamma}$ distribution around 750~GeV and the theoretical prediction from the toy model with various choices of the parameters. Clearly, if the mass splitting is small and the widths are large, as depicted in Fig.~\ref{fig:fitting}(f), there will be a hard time for the experimentalists to discriminate this model from models with only one single fat scalar. In other cases where either the two scalars are narrow or the mass splitting is large enough, future experimental data with finer binning will be able to tell us whether there are indeed two nearly degenerate resonances in this region.

\section{Dark matter connection}

\begin{figure*}
\includegraphics[width=0.3\linewidth]{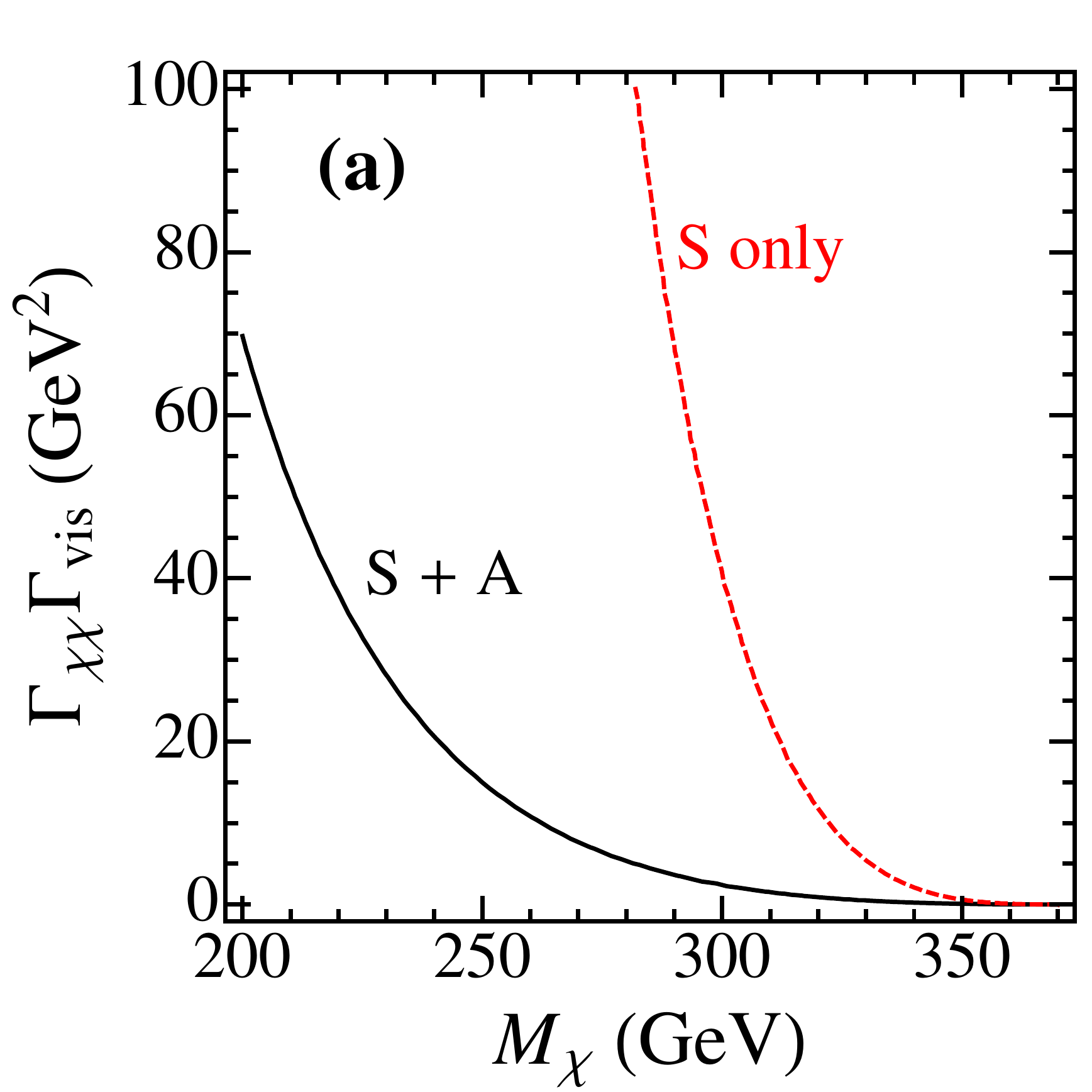}
\includegraphics[width=0.3\linewidth]{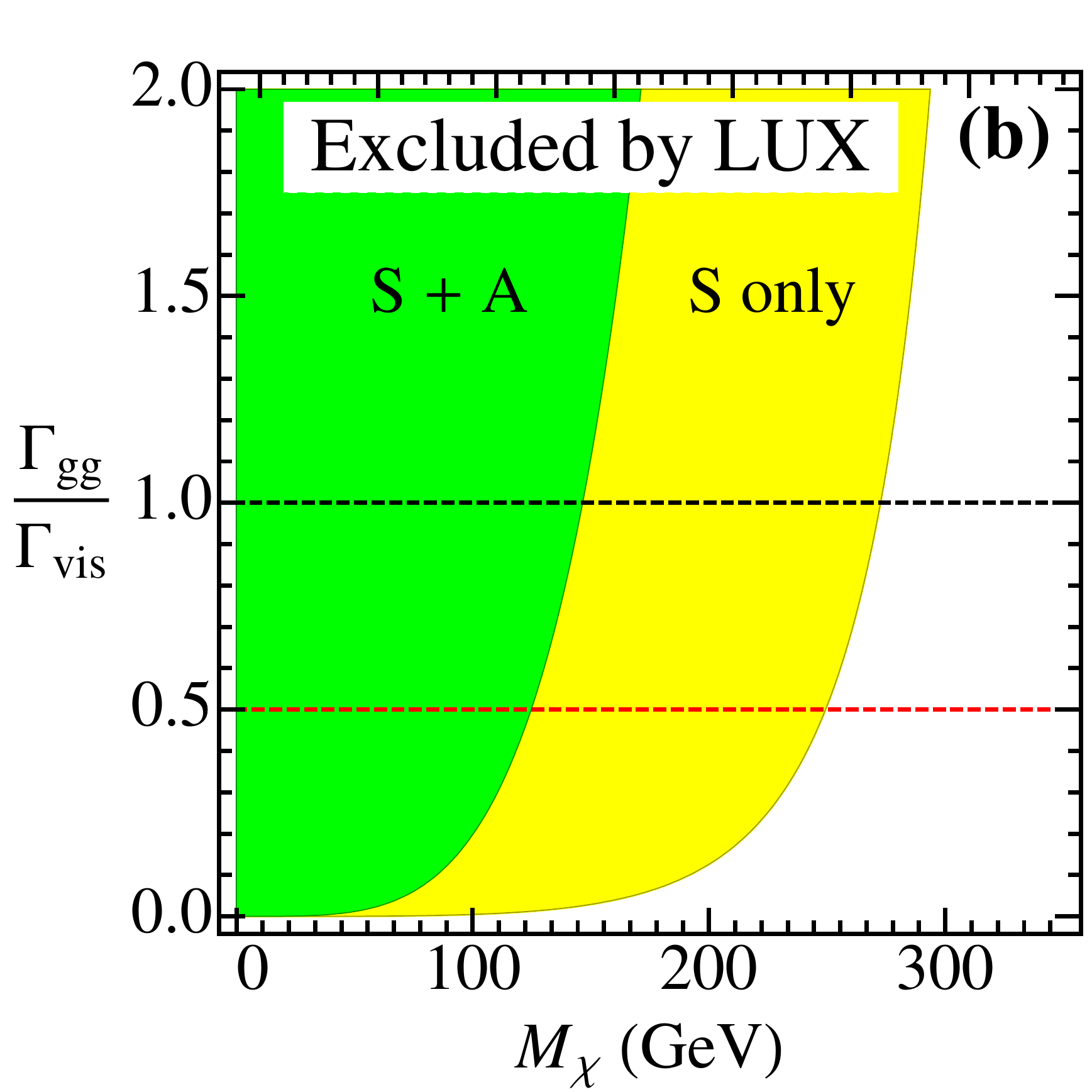}
\includegraphics[width=0.3\linewidth]{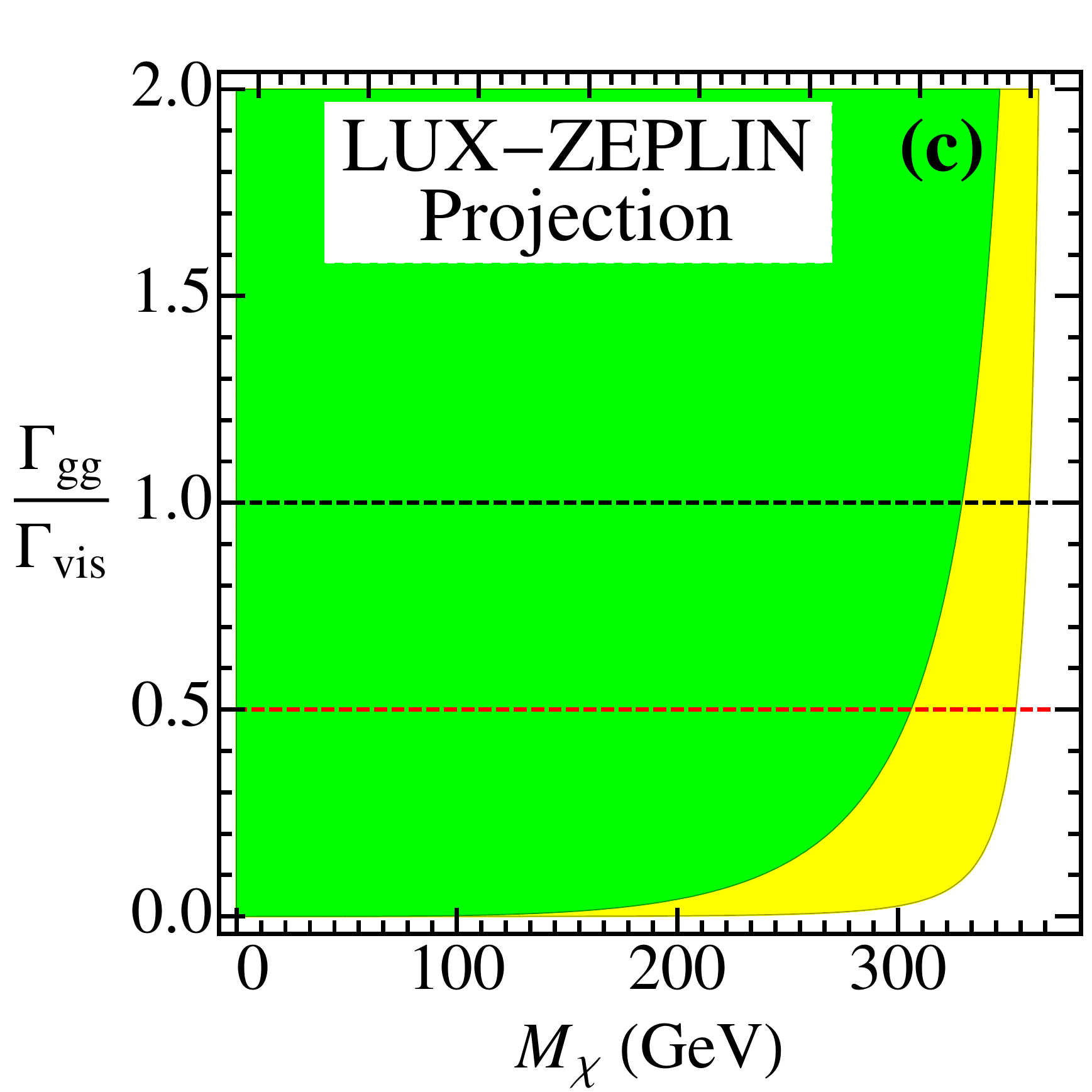}
\caption{(a) The needed $\Gamma_{\chi\chi}\Gamma_{\rm vis}$ to explain the relic abundance as a function of $M_{\chi}$; (b) the allowed parameter space of $\Gamma_{gg}/\Gamma_{\rm vis}$ by the LUX SI detection; (c) the LUX-ZEPLIN projection of the SI detection.}
\label{fig:relic}
\end{figure*}

While the assumption that $S$ and $A$ only decay into gauge bosons giving a very small total width is perfectly compatible with the experimental data in the two resonance model, it is still interesting to ask whether a large width is possible. A large width from decaying into other visible particles such as the SM Higgs bosons or the top quarks is in potential conflict with current experimental constraints. It is therefore often assumed that the large width comes from decaying into invisible particles such as the dark matter. In this section we assess such a possibility by coupling the complex scalar to a Majorana dark fermion $\chi$ with the Lagrangian
\begin{align}
  \mathcal{L}_{\text{dark}} = g_D^S S \bar{\chi} \chi + i g_D^A A \bar{\chi} \gamma^5 \chi \, .
\end{align}
The partial widths of $S$ and $A$ to a pair of dark fermions are given by
\begin{align}
  \Gamma^S_{\chi\chi} = \frac{\left( g_D^S \right)^2}{4\pi} M_S \beta_S^3 \, , \quad   \Gamma^A_{\chi\chi} = \frac{\left( g_D^A \right)^2}{4\pi} M_A \beta_A \end{align}
 where $\beta_S = \sqrt{ 1 - 4M_\chi^2/M_S^2 }$ and $\beta_A = \sqrt{ 1 - 4M_\chi^2/M_A^2 }$ with $M_\chi$ being the mass of the dark fermion. We will again make a simplifying assumption 
 \begin{align}
 \label{eq:assumptions2}
 \Gamma^S_{\chi\chi} = \Gamma^A_{\chi\chi} = \Gamma_{\chi\chi}, \qquad \Gamma^S=\Gamma^A =\Gamma,
 \end{align}
where $\Gamma^{i} = \Gamma_{gg}^{i} + \Gamma^{i}_{\gamma\gamma} +\Gamma^{i}_{WW}+\Gamma^{i}_{ZZ}+\Gamma^{i}_{Z\gamma}+ \Gamma^{i}_{\chi\chi}$ with $i=S/A$.  Furthermore, since the small mass splitting between the CP even and the CP odd components has little effects on the dark matter constraints, we will simply take $M_S = M_A = M$.

Such a simple model can be projected to several dark matter related observables: the dark matter relic abundance, the monojet cross section at the LHC, as well as the direct and indirect detections. 

We begin with the monojet searches. Both the ATLAS and CMS collaborations have performed searches for events with a high-$p_T$ jet and large missing transverse energy \cite{Khachatryan:2014rra, Aad:2015zva}. Upper limits on new physics cross sections were obtained which can be translated to upper limits on $\Gamma_{gg}\Gamma_{\chi\chi}/\Gamma$ in our model. 
We perform a simulation of the monojet production process using \texttt{MadGraph5} \cite{Alwall:2014hca} with model files generated by \texttt{FeynRules} \cite{Alloul:2013bka}, \texttt{Pythia} \cite{Sjostrand:2006za} and \texttt{Delphes} \cite{deFavereau:2013fsa}, and find that the most stringent constraint comes from the CMS data with $\slashed{E}_T > \unit{450}{\GeV}$, which leads to an upper bound $\Gamma_{gg} \Gamma_{\chi\chi}/\Gamma< \unit{0.13}{\GeV}$. In the following we will apply the conservative constraint $\Gamma_{gg} \Gamma_{\chi\chi}/\Gamma< \unit{0.2}{\GeV}$.

We now turn to the relic abundance. The annihilation rates of a pair of the dark fermions into a pair of SM particles ($x$ and $y$) in the non-relativistic limit are 
\begin{align}
  \sigma^S_{xy} v &= 64\pi \left( \frac{M_\chi}{M} \right)^4 \frac{\Gamma_{\chi\chi}\Gamma_{xy}}{M^4} \, \frac{v^2}{\beta^3 \left( \beta^4 + \Gamma^2/M^2 \right)} \, ,
  \\
  \sigma^A_{xy} v &= 256\pi \left( \frac{M_\chi}{M} \right)^4 \frac{\Gamma_{\chi\chi}\Gamma_{xy}}{M^4} \, \frac{1}{\beta \left( \beta^4 + \Gamma^2/M^2 \right)} \, ,  
\end{align}
where $v$ is the relative velocity between the two dark fermions, and $\beta = \sqrt{1-4M_\chi^2/M^2}$. Note that the annihilation rate through the CP odd mediator is enhanced by a factor of $4\beta^2/v^2$ with respect to the CP even case. Hence, the dark matter will mainly annihilate through the $A$ scalar into the SM particles. To explain the current relic abundance measured by the Planck experiment~\cite{Ade:2015xua}, $\Omega h^2=0.1186\pm 0.0020$, the thermal averaged annihilation cross section is  approximately
\begin{align}
\left<\sigma v\right>_{\rm Relic}\equiv \Braket{\sum_{xy} \left( \sigma^S_{xy}+\sigma^A_{xy} \right) v} \approx 0.83~{\rm pb}.
\end{align} 
Fixing $M=\unit{750}{\GeV}$, the required total annihilation rate determines a relation between the two parameters $M_\chi$ and $\Gamma_{\chi\chi}\Gamma_{\rm vis}$ in the parameter space of interests to us. We show such a correlation as the black solid curve in Fig.~\ref{fig:relic}(a). Here $\Gamma_{\rm vis}\equiv \sum_{xy} \Gamma_{xy}$ with $x$ and $y$ being SM particles. If only a CP-even mediator is present, a larger $\Gamma_{\chi\chi}\Gamma_{\rm vis}$ is needed to overcome the $p$-wave suppression, as demonstrated by the red dashed curve in Fig.~\ref{fig:relic}(a).

We further consider the constraints coming from the direct detection of the dark matter. The cross section for the spin-independent (SI) elastic scattering between a dark fermion and a nucleon is dominated by the exchange of the scalar component $S$, and can be obtained from the formulas given in Ref.~\cite{Berlin:2015ymu}. Ignoring the slight difference between the proton and the neutron, the cross section can be written in terms of the partial widths as
\begin{align}
  \sigma_{\text{SI}} = 8\pi \left( \frac{8\pi}{9\alpha_s} \right)^2 \frac{m_n^4}{M^4} \, \frac{\Gamma_{\chi\chi}\Gamma_{gg}}{M^4\beta^3} \, f_{TG}^2 \, ,
\end{align}
where the nucleon mass $m_n \approx \unit{0.94}{\GeV}$, the gluon fraction $f_{TG} \approx 0.94$ \cite{Berlin:2015ymu}. The most stringent limits on the cross sections for a dark matter mass between \unit{50}{\GeV} and \unit{370}{\GeV} come from the LUX experiment \cite{Akerib:2013tjd, Akerib:2015rjg}. 
We use the tight constraint from the relic abundance as a reference to study the parameter spaces of dark matter direct and indirect detections.  We compare the spin-independent scattering cross section to the relic abundance constraint, which gives rise to
\begin{align}
\frac{\sigma_{\rm SI}}{\sigma_{\rm Relic}} = \frac{1}{32}\left(\frac{8\pi}{9\alpha_S}f_{TG}\right)^2 \frac{m_n^2}{M_\chi^2} \frac{\beta^4 + \frac{\Gamma^2}{M^2}}{\beta^2+\frac{1}{4}\left<v^2\right>} \frac{\Gamma_{gg}}{\Gamma_{\rm vis}} \, .
\end{align}
Figure~\ref{fig:relic}(b) displays the allowed ratio $\Gamma_{gg}/\Gamma_{\rm vis}$ as a function of $M_\chi$. The shaded region is excluded by the LUX experiments~\cite{Akerib:2013tjd,Akerib:2015rjg}: (green) one scalar and one pseudoscalar mediators, (yellow) one CP-even scalar mediator. In the region between the black and red dashed curves, $\Gamma_{gg}>\Gamma_{\rm vis}/2$ while in the region below the red dashed curve $\Gamma_{gg}<\Gamma_{\rm vis}/2$. Adding in a CP-odd scalar mediator relaxes the LUX constraint on the case of only one CP-even scalar mediator. In Fig.~\ref{fig:relic}(c), we also plot the projection of LUX-ZEPLIN (LZ)~\cite{Akerib:2015cja} which is about two orders of magnitude smaller than the current LUX bounds. A vast of parameter space would be ruled out if no dark matter signals were observed in the LUX-ZEPLIN experiment; for example, $M_\chi$ needs to be larger than $\sim 150~{\rm GeV}$.

We will also consider indirect detections of the dark matter candidate, specifically focusing on the cosmic gamma ray constraints from the Fermi-LAT experiments~\cite{Ackermann:2013uma,Ackermann:2015lka}. Before discussing the gamma-ray constraints we summarize all the assumptions and constraints undertaken in our parameter scans as follows:\\
1) Scalar widths: 
\begin{align}
\label{eq:assumptions3}
&\Gamma=\Gamma_{\rm vis} + \Gamma_{\chi\chi},\nonumber\\
&\Gamma_{\rm vis}=\Gamma_{gg}+\Gamma_{\gamma\gamma}+\Gamma_{WW}+\Gamma_{ZZ}+\Gamma_{Z\gamma},
\end{align}
where $\Gamma_{\chi\chi}\leq 40~{\rm GeV}$ and $\Gamma_{\rm vis}\leq 10~{\rm GeV}$ are required to expedite the numerical scan; \\
2) Diphoton production~\cite{ATLAS:2015xxx,CMS:2015dxe}: 
\begin{align}
3~{\rm fb}<\sigma_{\gamma\gamma}\approx \frac{\Gamma_{gg}\Gamma_{\gamma\gamma}}{\Gamma}\times 19500~{\rm fb}\cdot{\rm GeV}^{-1}\leq 13~{\rm fb}; \nonumber
\end{align}
3) Dijet constraints~\cite{Aad:2014aqa}: 
\begin{align}
\frac{\Gamma_{gg}^2}{\Gamma}\leq \unit{0.6}{\GeV};
\end{align}
4) Diboson constraints~\cite{Aad:2015kna,Khachatryan:2015cwa,Aad:2015kna,Aad:2015agg}:
\begin{align}
\frac{\Gamma_{gg}\Gamma_{Z\gamma}}{\Gamma} &\leq 0.96\times 10^{-3}~{\rm GeV},\nonumber\\
\frac{\Gamma_{gg}\Gamma_{ZZ}}{\Gamma} &\leq 2.9\times 10^{-3}~{\rm GeV},\nonumber\\
\frac{\Gamma_{gg}\Gamma_{WW}}{\Gamma} &\leq 9.6\times 10^{-3}~{\rm GeV};
\end{align}
5) Mono-jet plus $\met$~\cite{Khachatryan:2014rra, Aad:2015zva}: 
\begin{align}
\frac{\Gamma_{gg}\Gamma_{\chi\chi}}{\Gamma}\leq 0.2~{\rm GeV};
\end{align}
6) Relic Abundance~\cite{Ade:2015xua}: 
\begin{align}
\left<\sigma v\right>_{\rm Relic} \approx 0.83~{\rm pb};
\end{align}
7) SI direct detection bounds and projection~\cite{Akerib:2013tjd,Akerib:2015rjg,Akerib:2015cja}: 
\begin{align}
&{\rm LUX:}  &&  \sigma_{\rm SI}\leq (1.0\text{--}1.5)~{\rm zb},\nonumber \\
&{\rm LZ~Projection:} && \sigma_{\rm SI}\leq (1.0\text{--}1.5)\times 10^{-2}~{\rm zb};
\end{align}
In the numerical scan we choose the following set of  independent parameters:
\begin{align}
M_\chi, ~\Gamma_{\chi\chi}, ~\Gamma_{gg}, ~\Gamma_{\gamma\gamma}, ~\Gamma.
\end{align}
Note that $\Gamma_{ZZ/WW/Z\gamma}$ can be uniquely determined from the 3 variables $\Gamma_{\rm vis}=\Gamma-\Gamma_{\chi\chi}$, $\Gamma_{gg}$ and $\Gamma_{\gamma\gamma}$ as all the partial widths to visible SM particles depend only on 3 operators.

\begin{figure}
\includegraphics[scale=0.4]{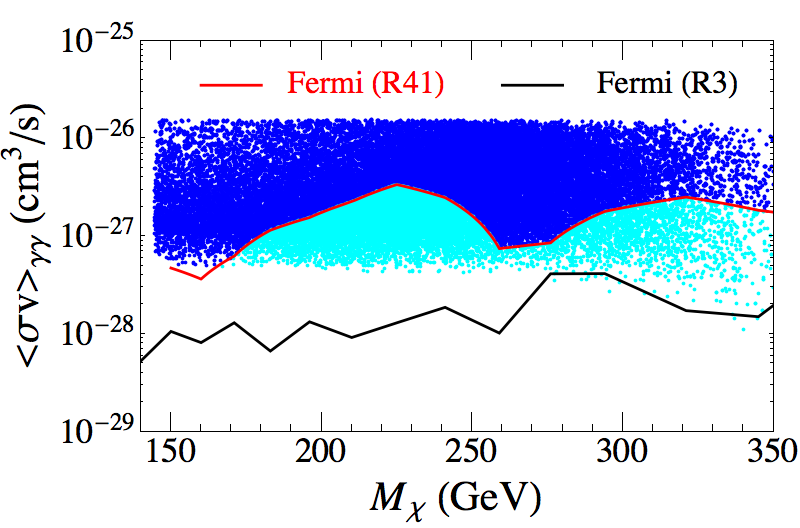}
\caption{The parameter space projected in the plane of $M_\chi$ and $\braket{\sigma v}_{\gamma\gamma}$. All the points satisfy the constraints from the collider searches, the dark matter relic abundance and the LUX bound on the spin-independent direct detection cross section. The black curve denotes the Fermi-LAT gamma-ray line limit from region R3 while the red curve represents the limit from region R41.}
\label{fig:fermi}
\end{figure}

Based on all the above constraints, we now proceed to take into account the cosmic gamma-ray constraints imposed by the Fermi-LAT experiment~\cite{Ackermann:2013uma,Ackermann:2015lka}. Figure~\ref{fig:fermi} displays the allowed parameter space in the plane of $M_\chi$ and $\braket{\sigma v}_{\gamma\gamma}$. The black curve denotes the Fermi gamma-ray line limit from the region R3 while the red curve represents the limit from the region R41. We first notice that almost the entire parameter space is excluded by the R3 limit. Therefore, if the R3 limit is indeed robust, the complex scalar responsible for the diphoton excess cannot be the mediator between the dark matter and the visible sector. In this case, we arrive at the conclusion that the two resonances must exhibit narrow widths and future LHC data in the diphoton channel will be able to distinguish the two peaks.

\begin{figure}[t!]
\includegraphics[scale=0.39]{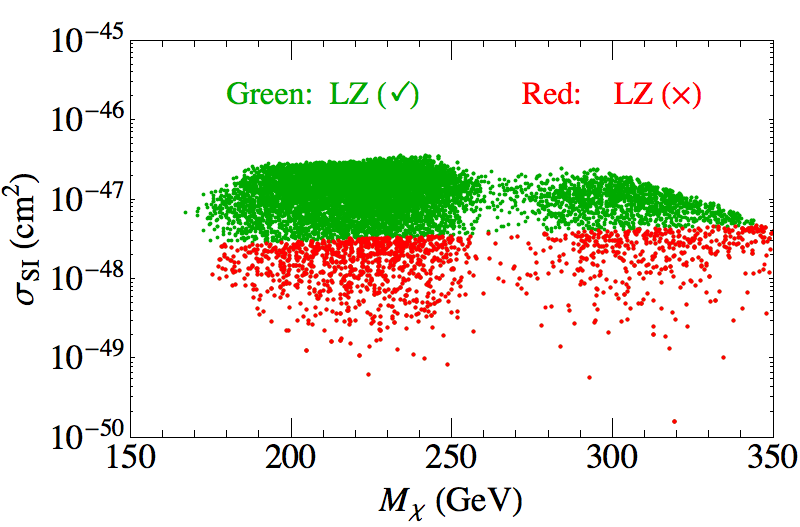}
\caption{The parameter space allowed by the collider searches, the dark matter relic abundance, the LUX direct detection bound, and the Fermi-LAT gamma-ray limit from region R41 in the $M_\chi$-$\sigma_{\rm SI}$ plane. The green points can be probed by the LZ experiment in the near future, while the red points are not testable by the foreseeable direct detection experiments.}
\label{fig:lz}
\end{figure}

Alternatively, we consider the possibility that there might be some unknown systematics in the R3 region and the gamma-ray constraints are less restrictive. We take for reference the constraints from the R41 region as an example. In Fig.~\ref{fig:fermi}, the blue points are excluded by the R41 data while the cyan points are allowed. Starting from the parameter space represented by the cyan points, we now proceed to investigate the possible signatures which may be detected in future experiments. In Fig.~\ref{fig:lz}, we demonstrate the potential of the LZ experiment to probe the model under consideration. The green points represent the parameter space which can be experimentally tested in the near future, while if the parameters lie in the region of the red points, foreseeable direct detection experiments will not have enough sensitivity. In the latter case, it is essential to study if there exist other possible collider signatures to supplement the diphoton one.

\begin{figure}[h!]
\includegraphics[scale=0.3]{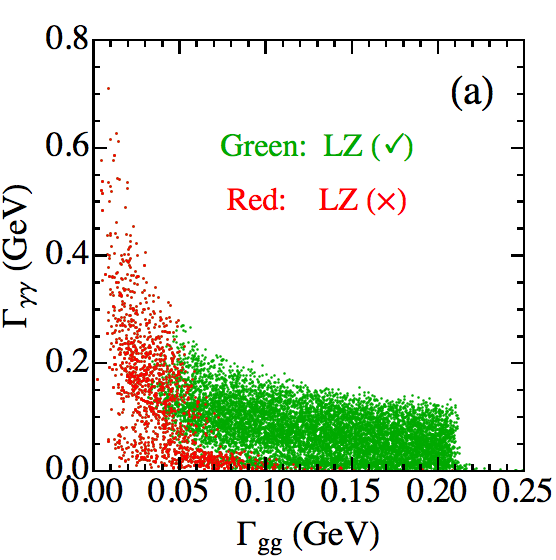}
\includegraphics[scale=0.3]{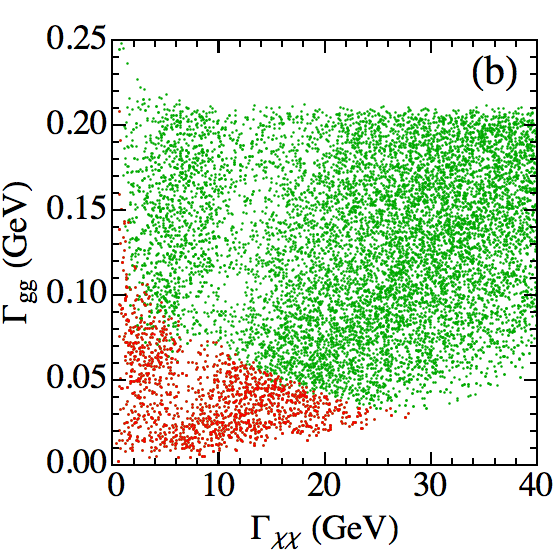}\\
\includegraphics[scale=0.3]{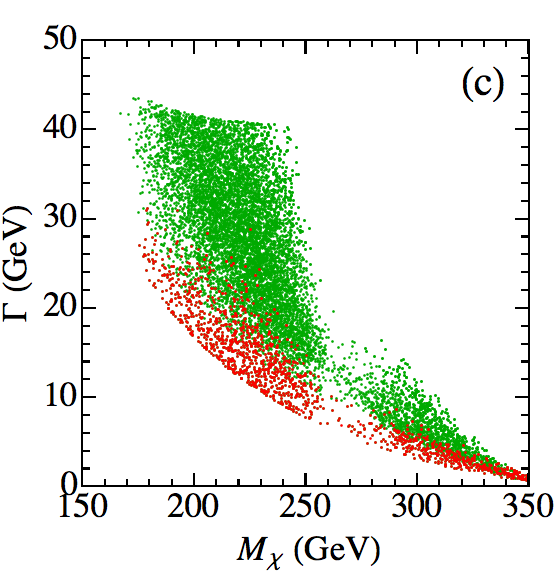}
\includegraphics[scale=0.3]{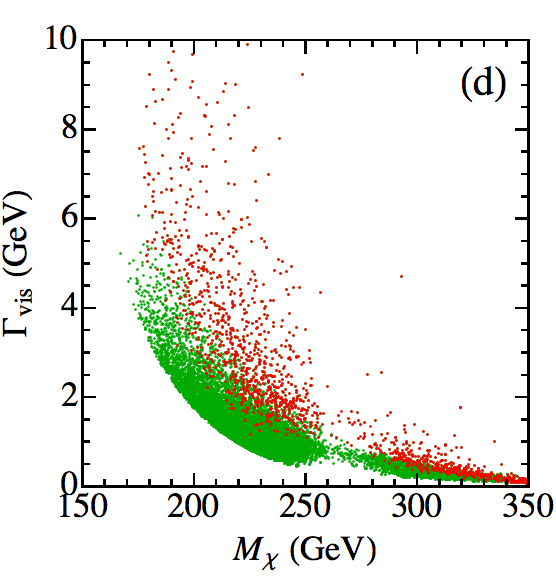}\\
\includegraphics[scale=0.5]{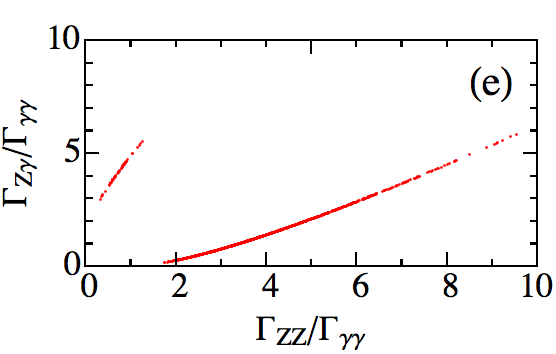}
\caption{The parameter space allowed by the collider searches, the dark matter relic abundance, the LUX direct detection bound, and the Fermi-LAT gamma-ray limit from region R41 in the $M_\chi$-$\sigma_{\rm SI}$ plane. The green points can be probed by the LZ experiment in the near future, while the red points are not testable by the foreseeable direct detection experiments.}
\label{fig:scan}
\end{figure}

In Fig.~\ref{fig:scan} we show the allowed ranges of a few parameters relevant for collider searches, where the green and red points have the same meanings as in Fig.~\ref{fig:lz}. Fig.~\ref{fig:scan}(a) displays the correlation between the partial widths to photons and to gluons, where the constraint $\Gamma_{gg} \lesssim \unit{0.2}{\GeV}$ mainly comes from the monojet search. If the future LZ experiment does not detect the dark matter, as the red points indicate, it is very likely that $\Gamma_{\gamma\gamma}$ is of the same order as or even larger than $\Gamma_{gg}$. This poses challenges to model building due to the smallness of the electromagnetic coupling compared to the strong coupling. One need to introduce a large amount of degrees of freedom which couple to photons but not to gluons, which is possible in new physics models. 

From Fig.~\ref{fig:scan}(b), one can see that in order to achieve a discovery of the dark matter at the LZ experiment, either $\Gamma_{gg}$ or $\Gamma_{\chi\chi}$ needs to be large. This can be expected since the direct detection cross section is most sensitive to the coupling of the scalar mediator to the dark matter and to gluons. On the other hand, if LZ does not see any dark matter, we may roughly arrive at the upper limits $\Gamma_{gg} \lesssim \unit{0.15}{\GeV}$ and $\Gamma_{\chi\chi} \lesssim \unit{28}{\GeV}$, while from Fig.~\ref{fig:scan}(c) we may obtain an upper limit on the total width, $\Gamma \lesssim \unit{30}{\GeV}$. However, in this case, one may see from Fig.~\ref{fig:scan}(d) that the total visible width $\Gamma_{\text{vis}}$ tends to have a large value if the dark matter is relatively light (say, e.g., $M\chi < \unit{250}{\GeV}$). This means that some of the visible modes besides the diphoton channel might be detectable at the LHC. In Fig.~\ref{fig:scan}(e) we plot the two ratios $\Gamma_{ZZ}/\Gamma_{\gamma\gamma}$ and $\Gamma_{Z\gamma}/\Gamma{\gamma\gamma}$, which can be translated to the production rate of the $ZZ$ and $Z\gamma$ final states through the complex mediator. Note that here we only take the red points which can evade the LZ experiment and we also demand $M_\chi < \unit{250}{\GeV}$. From the plot we see that either $\Gamma_{ZZ}/\Gamma_{\gamma\gamma}$ or $\Gamma_{Z\gamma}/\Gamma{\gamma\gamma}$ needs to be large, meaning that if the diphoton signal is confirmed in the future data collected at the LHC, either the $ZZ$ channel or the $Z\gamma$ channel may also be observable.

Finally, if the dark matter mass is close to half the mass of the scalar mediator, say $M_\chi > \unit{300}{\GeV}$, one relies on the Breit-Wigner resonance effect to produce the correct relic density, and all the couplings of the scalar mediator may be relatively small. In this case no signals other than the diphoton channel can be observed at the LHC. Of course, all the above analyses are partly based on the simple assumptions we made in Eqs.~(\ref{eq:assumptions1}, \ref{eq:assumptions2}, \ref{eq:assumptions3}). If some of those assumptions are relaxed, one may expect certain level of relaxation of the conclusions we arrived at.

\section{Conclusion}

The recently observed broad excess in the diphoton invariant mass spectrum around 750 GeV is usually interpreted as a singlet scalar with a large width in new physics models. However, it is difficult in an UV completed model to generate such a large width for a scalar and in the same time a large enough branching ratio to photons. It is often assumed that the scalar decays significantly into a pair of dark matter candidates. However, a strong tension is found between the diphoton production and the null results of dark matter detection experiments.
In this work we demonstrate that the diphoton resonance could be composed by two scalar mediators. The broad excess around 750 GeV can be easily explained as the overlap of the two scalar resonances as long as the two scalars exhibit a proper mass splitting. Our fitting shows that the two scalars can be either narrow or wide, which cannot be determined from the current data yet. In the case that both scalars are narrow, there is no need to introduce the decay mode of the scalar into dark matter candidates. This alleviates the tension between the large width of the scalar and dark matter detection data. 

Even though two narrow scalars is enough to explain the diphoton excess, we nevertheless consider the possibility that the two scalars may have larger widths by interacting with light dark matter candidates. We introduce a toy model consisting a SM singlet complex scalar ($S=H+iA$) and a Majorana fermionic dark matter ($\chi$). The interactions of the scalars with the SM gauge bosons are described by three dimension-5 operators. We then explore how wide the two scalars could be in this scenario. For simplification we assume that both scalars share similar masses and partial widths. After taking into account various constraints from the collider and dark matter experiments, we find that the cosmic gamma-ray line limits from the Fermi-LAT experiment constrain the parameter space significantly. Almost the entire parameter space is ruled out by the gamma-ray line limit from the region R3. If the R3 limit stands robustly, then the complex scalar responsible for the diphoton anomaly cannot be the mediator between the dark matter and the visible sector. However, the R3 limit might be relaxed if there are some unknown systematics in the R3 region. We then consider the R41 limit which excludes the parameter space of $M_\chi\lesssim 170~{\rm GeV}$. A part of the remaining parameter space could be probed in the future LUX-ZEPLIN experiment if the product $\Gamma_{gg} \Gamma_{\chi\chi}$ is large enough. On the other hand, if the LZ experiment does not see any dark matter signal, we may roughly arrive at the upper limits on the scalar partial widths $\Gamma_{gg}\lesssim 0.15~{\rm GeV}$, $\Gamma_{\chi\chi}\lesssim 28~{\rm GeV}$ and total width $\Gamma\lesssim 30~{\rm GeV}$. 

So far the diphoton excess may still turn out to be a statistical fluctuation. One cannot draw any definitive conclusion on its origin based on the current experimental data. If the diphoton anomaly is confirmed in the future LHC data, then measuring its property will shed lights on new physics beyond the SM. Future experimental data with finer binning might be able to tell us whether there are indeed two nearly degenerate resonances in this region.

\begin{acknowledgments}
This work was supported in part by the National Natural Science Foundation of China under Grant No. 11275009 and 11575004.
\end{acknowledgments}

\bibliographystyle{apsrev}
\bibliography{reference}

\end{document}